\documentclass[11pt,twoside]{article}
\usepackage{asp2006}
\usepackage{epsf}
\usepackage{lscape}
\begin{document}
\title{
Morphological Transformations of Galaxies in the A901/02 Supercluster 
from STAGES
}
\vspace{-0.2cm} \author{A.L. Heiderman (UT Austin), S. Jogee (UT
Austin), D.J. Bacon (Portsmouth), M.L. Balogh (Waterloo), M. Barden
(Innsbruck), F.D. Barazza (EPFL), E.F. Bell (MPIA), A. B\"ohm (AIP),
J.A.R. Caldwell (UT Austin), M.E. Gray (Nottingham), B. H\"au\ss ler
(Nottingham), C. Heymans (UBC, IAP), K. Jahnke (MPIA), E. van Kampen
(Innsbruck), S. Koposov (MPIA), K. Lane (Nottingham), D.H. McIntosh
(UMass), K. Meisenheimer (MPIA), C. Y. Peng (NRC-HIA), H.-W. Rix
(MPIA), S.F. Sanchez (CAHA), R. Somerville (MPIA), A.N. Taylor (SUPA),
L. Wisotzki (AIP), C. Wolf (Oxford), \& X. Zheng (PMO)}
\vspace{-0.4cm}

\begin{abstract}
We present a study of galaxies in the Abell 901/902 Supercluster at
$z\sim$~0.165, based on $HST$ ACS F606W, COMBO-17, $Spitzer$
24$\micron$, XMM-Newton X-ray, and gravitational lensing maps, as part
of the STAGES survey.  We characterize galaxies with strong
externally-triggered morphological distortions and normal relatively
undisturbed galaxies, using visual classification and quantitative CAS
parameters. We compare normal and distorted galaxies in terms of their
frequency, distribution within the cluster, star formation properties,
and relationship to dark matter (DM) or surface mass density, and
intra-cluster medium (ICM) density. We revisit the morphology density
relation, which postulates a higher fraction of early type galaxies in
dense environments, by considering separately galaxies with a low
bulge-to-disk ($B/D$) ratio and a low gas content as these two
parameters may not be correlated in clusters. We report here on our
preliminary analysis.
\end{abstract}

\vspace{-0.3 cm}
\section{Introduction}
The systematic quest to understand how galaxies evolve as a function
of epoch and environment remains in its infancy.  Galaxies in cluster
environments may differ from field galaxies due to high initial
densities leading to early collapse.  Furthermore, the relative
importance of galaxy-galaxy interactions (e.g., galaxy harassment,
tidal interactions, minor mergers, and major mergers) and galaxy-ICM
interactions (e.g, ram pressure stripping and compression) are likely
to differ between cluster and field environments due to the different
number density of galaxies, galaxy velocity dispersions, ICM density,
and DM density.

In order to constrain how galaxies evolve in cluster environments, we
present a study based on the STAGES survey of the Abell 901/902
supercluster (Gray et al. 2008).  The survey covers 0.5 $\times$ 0.5
degrees on the sky, and includes high resolution ($0.1\arcsec$,
corresponding to 280 pc at $z\sim$~0.165\footnote{We assume in this
paper a flat cosmology with $\Omega_M = 1 - \Omega_{\Lambda} = 0.3$
and $H_{\rm 0}$ =70~km~s$^{-1}$~Mpc$^{-1}$.})  $HST$ ACS F606W images,
along with COMBO-17, $Spitzer$ 24$\micron$, XMM-Newton X-ray data, and
gravitational lensing maps (Gray et al.  2002; Heymans et
al. 2008). The STAGES survey is complemented with accurate
spectrophotometric redshifts with errors $\delta_{\rm z}$/(1 +
$z$)$\sim$~0.02 down to $R_{\rm Vega}$=24 from the COMBO-17 survey
(Wolf et al. 2004), and stellar masses (Borch et al. 2006). Star
formation rates (SFRs) are derived from the COMBO-17 UV and $Spitzer$
24$\micron$ data (Bell et al. 2007).

The supercluster sample of 2309 galaxies covers a broad range of
luminosities, encompassing both dwarf and larger galaxies (E to Sd).
We present many of our results separately for bright (M$_{\rm V} \leq
$ -18; 798 galaxies), and faint galaxies (-18 $< \rm M_{\rm V} \leq
-15.5$; 1286 galaxies).  This is because the morphological
characterization is less robust for small dwarf galaxies, than for
normal galaxies, due to surface brightness, spatial resolution (280
pc), and contamination from field galaxies.
\vspace{-0.2cm}
\section{Methodology and Preliminary Results}
\indent Using the CAS code (Conselice et al. 2000), the concentration
(C), asymmetry (A), and clumpiness (S) parameters were derived from
the $HST$ ACS F606W images.  Given that the CAS merger criteria (A$>$S
and A$>$0.35) tend to capture only a fraction of interacting/merging
galaxies (e.g., Conselice 2006; Jogee et al. 2007), we also
characterize the morphological properties via visual classification.
In order to identify galaxies that are distorted due to a recent
merger or tidal interaction, we take special care to classify galaxies
into three distinct visual classes: (1) Galaxies with {\it
externally-triggered} distortions: These distortions, triggered by
tidal interactions or mergers, include double or multiple nuclei
inside a common body, tidal tails, arcs, shells, ripples, or tidal
debris in body of galaxy, warps, offset rings, and extremely
asymmetric SF or spiral arms on one side of the disk.  Galaxies are
classified as strongly or weakly distorted, according to whether the
distortions occupy a large or small fraction of the total light.  (2)
Galaxies with {\it internally-triggered} asymmetries (classified as
Irr-1): Internally-triggered asymmetries are due to stochastic
star-forming regions or the low ratio of ordered to random motions,
common in irregular galaxies.  These asymmetries tend to be correlated
on scales of a few hundred parsecs, rather than on scales of a few
kpc.  (3) Relatively {\it undistorted symmetric} galaxies (classified
as Normal).

\vspace{0.15 cm} The assumed correlation implicit in the Hubble
sequence, between a high $B/D$ and a low gas and dust content, can
often fail in cluster environments, where systems of low $B/D$ may
have smooth featureless disks (e.g., Koopmann \& Kenney 1998).  This
begs the question of whether the classical morphology-density
relation, which claims a larger fraction of early-type galaxies in
dense environments, is driven by a larger fraction of gas-poor
galaxies in such environments, or a larger fraction of bulge-dominated
systems, or both.  In order to address this question, we visually
classify the clumpiness and $B/D$ ratio of galaxies separately,
classifying galaxies into five categories: pure bulge (pB), bulge plus
disk (B+D) clumpy or smooth, and pure disk (pD) clumpy or smooth.
\vspace{0.15 cm}
We present below extracts  of our preliminary findings.
 
\vspace{0.3 cm} {\it 1. Fraction of strongly distorted galaxies:}
Visual classification shows that $\sim$3$\%$ and $\sim$0.4$\%$ of
bright (M$_{\rm V} \leq$ -18) and faint (-18 $< \rm M_{\rm V} \leq
-15.5$) galaxies are strongly distorted, respectively.  (The results
for faint galaxies are highly uncertain.).  The distortion fraction
for bright galaxies in Abell 901/902 is lower than the values of 7\%
to 9\% reported in the field for bright (M$_{\rm B} \le$ -20; Lotz et
al. 2007) or massive ($M \ge$~$2.5 \times 10^{10}$ $M_{\odot}$; Jogee
et al. 2007) galaxies.

\vspace{0.0 cm}
{\it 2. CAS  recovery rate:}  
The CAS merger criteria (A$>$S and A$>$0.35) capture
54\% and 40\% of the strongly distorted galaxies in the bright 
and faint samples, respectively.

\vspace{0.0 cm} {\it 3. Distribution of strongly distorted vs normal
galaxies:} Most strongly distorted galaxies lie outside the cluster
cores, avoid the peaks in DM surface mass density ($\kappa \geq$ 0.1)
and ICM density (Fig. 1A). These results are consistent with high
galaxy velocity dispersions in the core being unfavorable to mergers
and strong tidal interactions. It may also be due, at least in part,
to the predominance in the core of gas-poor systems (see point 6
below; Fig. 1B), which tend to show shorter-lived tidal signatures.
 
\vspace{0.0 cm} {\it 4. SFR of strongly distorted vs normal galaxies:}
The UV-based SFR ranges primarily from 0.001 to 15 $M_{\sun}$
yr$^{-1}$ and rises with stellar mass (Fig. 1D).  The average
SFR$_{\rm UV}$ is enhanced only by a modest factor of $\sim$~4 in
strongly distorted galaxies compared to normal galaxies (Fig. 1C). A
similar result is reported for distorted galaxies in field galaxies at
$z\sim$~0.2 (e.g., Jogee et al. 2007).  For 12\% of the 798 bright
galaxies having $Spitzer$ 24$\micron$ detection, the UV+IR-based SFR
ranges primarily from 0.01 to 50 $M_{\sun}$ yr$^{-1}$, and the median
ratio of (SFR$_{\rm UV+IR}$/SFR$_{\rm UV}$) is $\sim$~3, indicating
significant amounts of obscured star formation.

\vspace{0.0 cm} {\it 5.  Specific SFR (SSFR) of strongly distorted vs
normal galaxies:} The SSFR or SFR per unit stellar mass is on average
lower at higher stellar mass, consistent with a large fractional
growth happening in lower mass galaxies at later times (Fig. 1D).
Only a modest enhancement in SSFR$_{\rm UV}$ is seen in distorted
galaxies compared to normal galaxies.

\vspace{0.0 cm}
{\it 6. The Morphology-Density relation in A901/02:} Smooth (gas-poor)
galaxies cluster in regions of high ICM density while clumpy
(gas-rich) systems dominate at low ICM densities (Fig 1B).  The ratio
of smooth (gas-poor) galaxies to very clumpy (gas-rich) galaxies
within the central 0.3 Mpc is (89$\%$:11$\%$), (75$\%$:25$\%$), and
(86$\%$:14$\%$), respectively, for A901a, A901b, and A902. These
values are comparable to the ratio of (80\%:20\%) for early-type to
late-type galaxies in the classical morphology-density relation
(Dressler 1980).  Future work will explore whether the ratio of
bulge-dominated galaxies to disk-dominated galaxies also shows a
similar trend with galaxy number density.

\vspace{-0.1 cm}
\acknowledgements
AH and SJ acknowledge support from NSF grant AST-0607748,
LTSA grant NAG5-13063, and HST-GO-10861 from STScI, which is
operated by AURA, Inc., for NASA, under NAS5-26555.

\setcounter{figure}{0}
\begin{figure}[!ht]
\plottwo{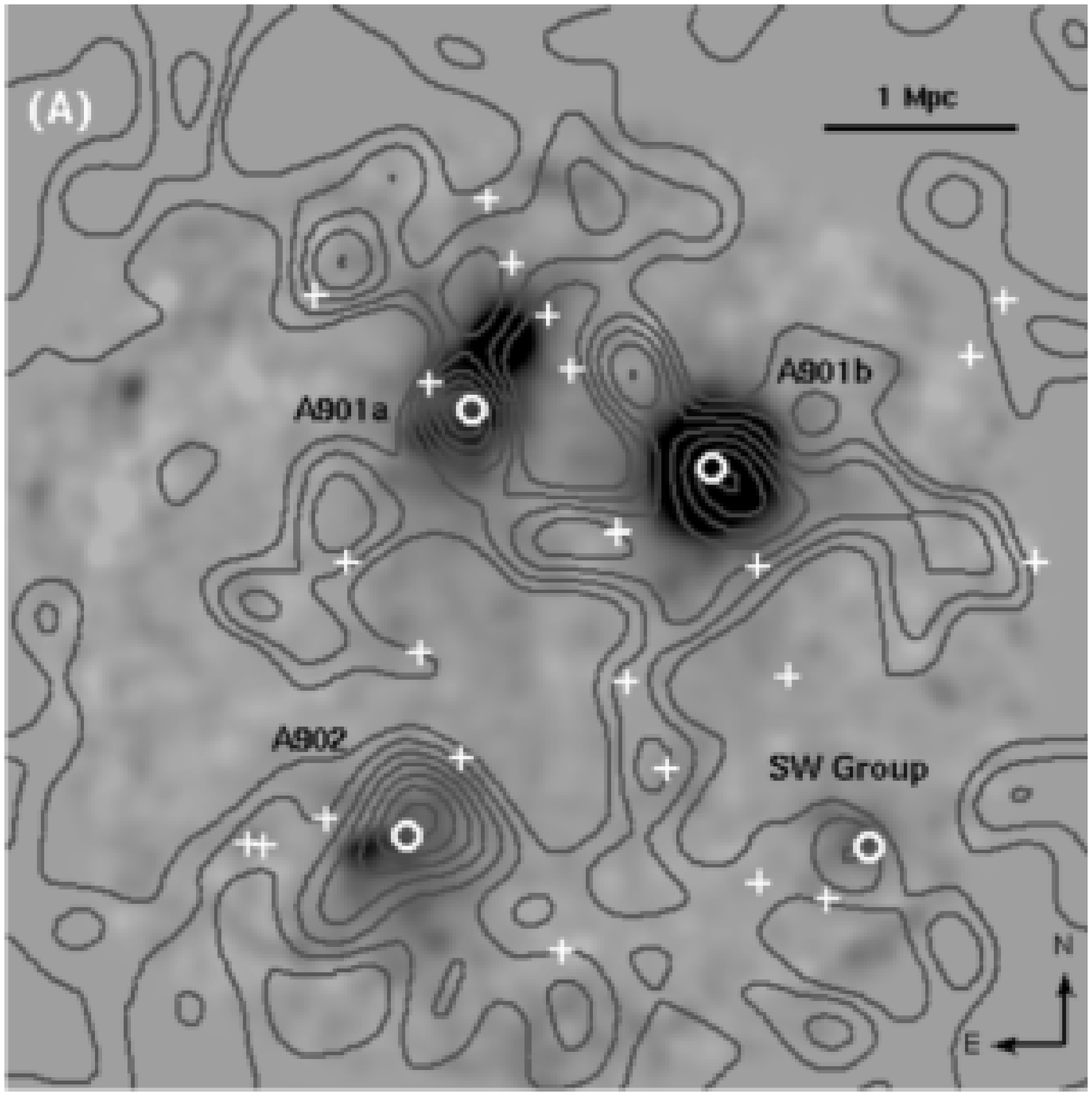}{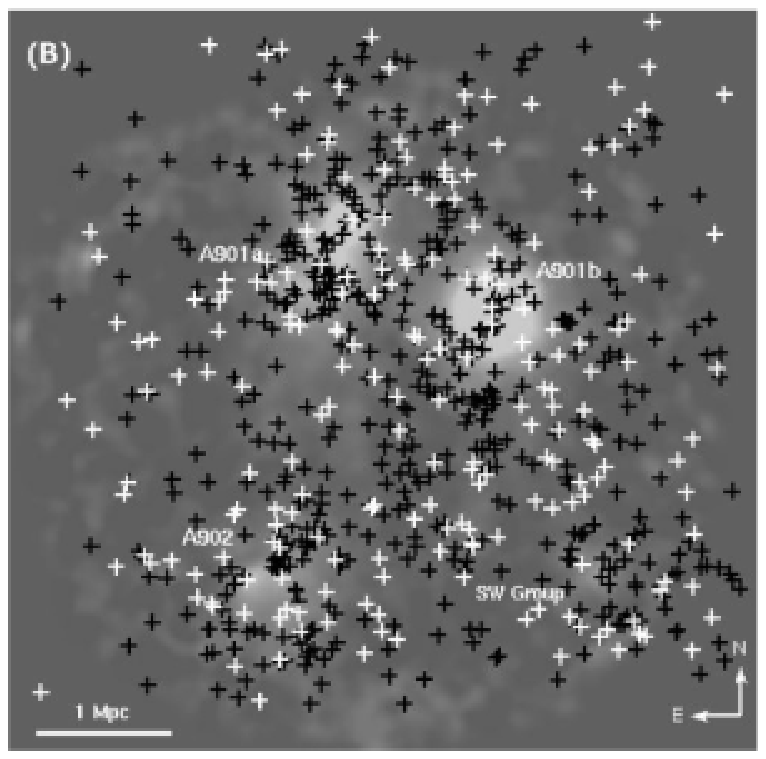}
\plottwo{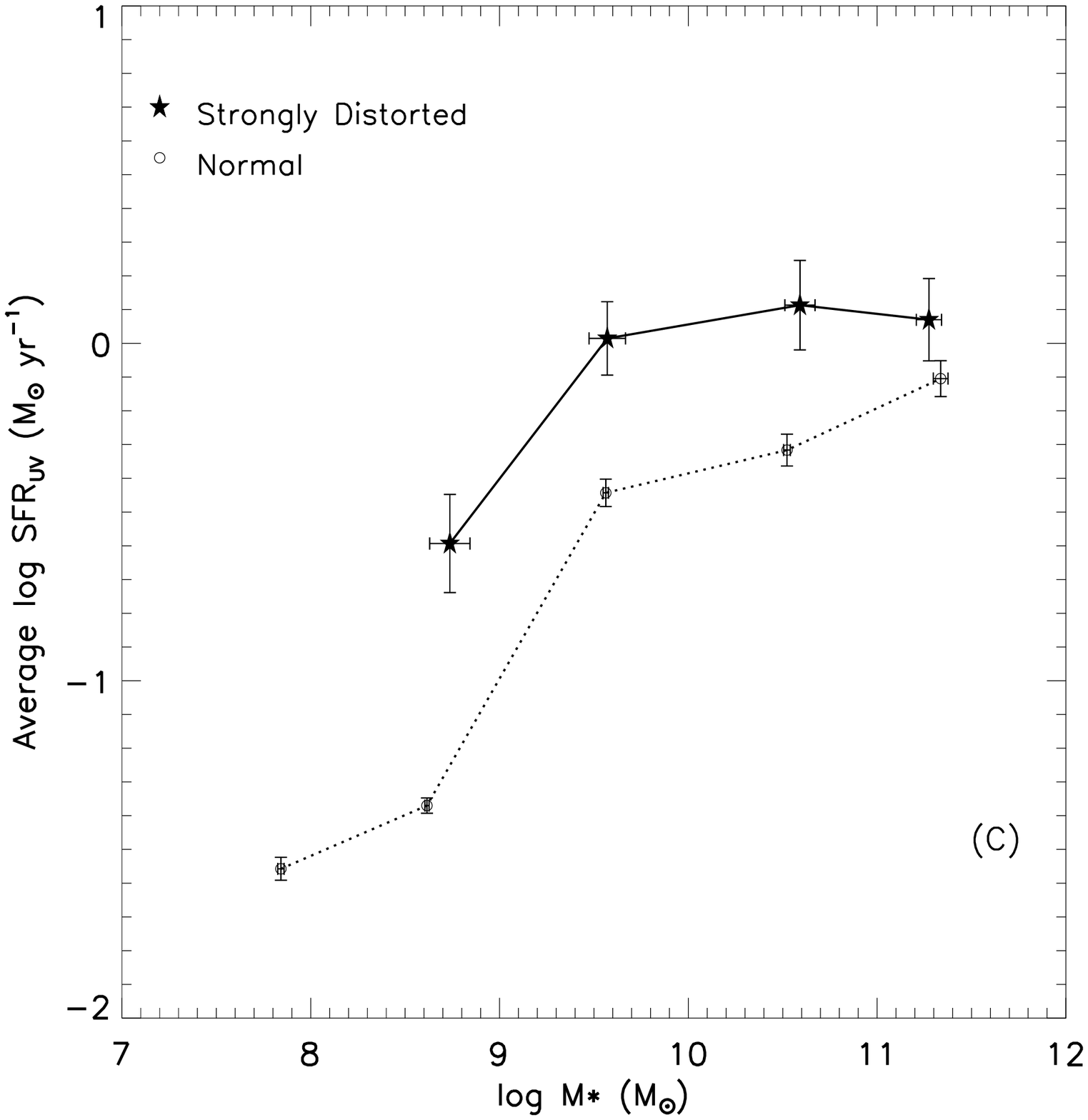}{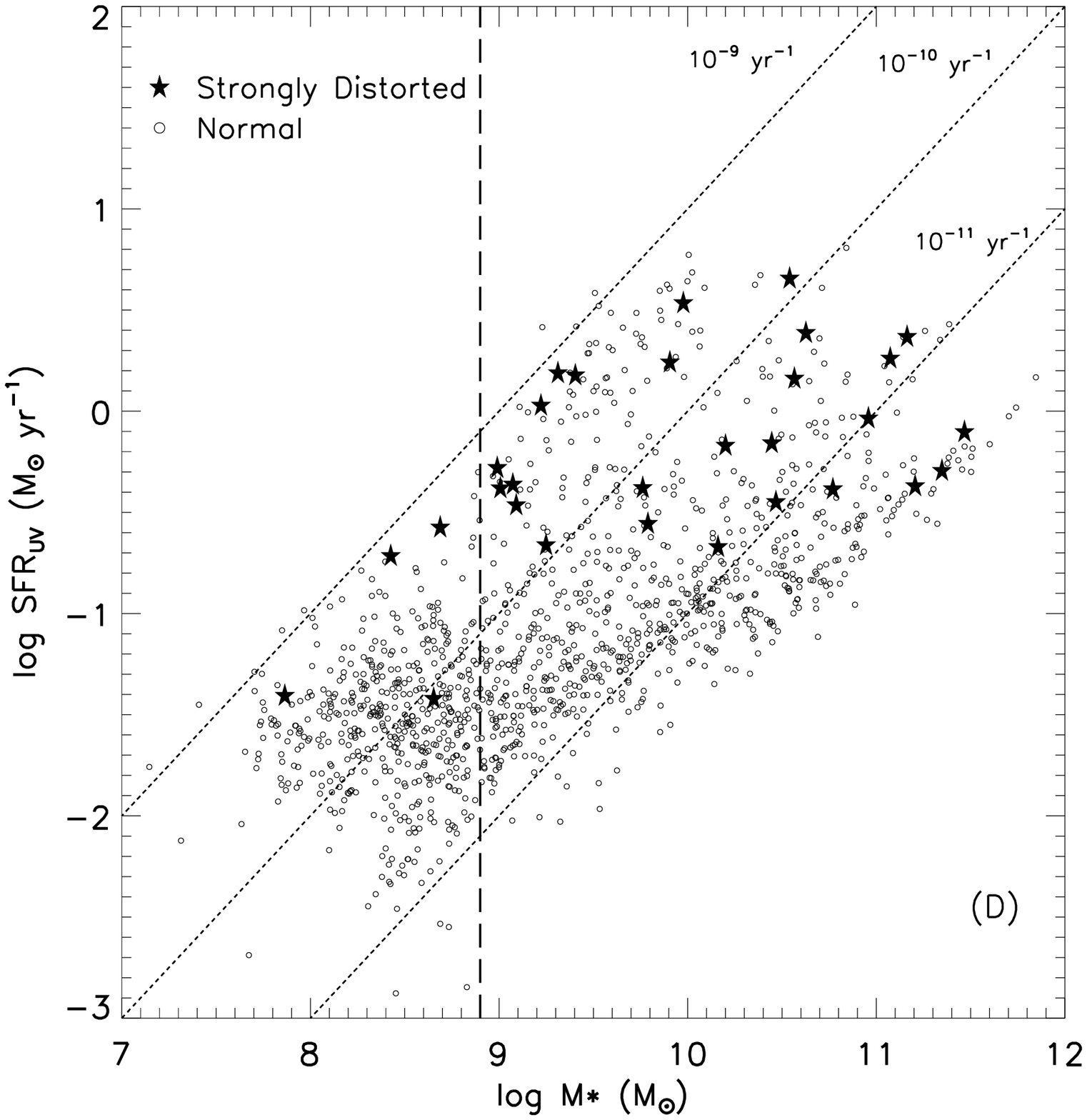}
\caption{ {\bf (A)} Strongly distorted galaxies (white crosses) are
overplotted on the ICM density (greyscale) and DM surface mass density
$\kappa$ (contours ranging from 0.2 to 0.12, in steps of 0.2). Most
strongly distorted galaxies lie outside region of high ICM density and
DM peaks ($\kappa \sim 0.1$).  {\bf (B)} We revisit the
morphology-density relation by overplotting the distribution of bright
(M$_{\rm V} \leq -18$) smooth (gas-poor) galaxies (black crosses) and
clumpy (gas-rich) galaxies (white crosses) on the ICM density.  Smooth
(gas-poor) galaxies populate the highest density regions.  {\bf (C)}
The average SFR$_{\rm UV}$ is plotted as a function of stellar
mass. It is enhanced by a modest factor of 4 in strongly distorted
galaxies compared to normal systems.  {\bf (D)} Strongly distorted
(black filled stars) and normal relatively undisturbed (open circles)
galaxies are plotted on the SFR$_{\rm UV}$ {\it versus} stellar mass
plane.  Loci of constant specific SFR$_{\rm UV}$ are marked in units
of yr$^{-1}$.  The dashed vertical line denotes the COMBO-17 mass
completeness limit for the red sequence. Note that SFR$_{\rm UV}$ for
red sequence galaxies are upper limits.  }
\end{figure}
\end{document}